\documentclass[preprint,amsmath,amssymb]{revtex4}

\usepackage{graphicx}
\usepackage{dcolumn}
\usepackage{bm}

\begin{document}


\title{Nanorod optical antennas for dipolar transitions}

\author{Tim H. Taminiau$^1$}
\email{Tim.Taminiau@icfo.es}
\author{Fernando D. Stefani$^2$}
\author{Niek F. van Hulst$^1$}
\altaffiliation{ICREA - Instituci\'{o} Catalana de Recerca i Estudis
Avan\c{c}ats, Spain.}
\affiliation{$^1$ICFO-Institut de Ciencies
Fotoniques,
Mediterranean Technology Park, 08860 Castelldefels (Barcelona), Spain.\\
$^2$Photonics and Optoelectronics Group,
Ludwig-Maximilians-Universit\"{a}t M\"{u}nchen and CeNS,
Amalienstrasse
54, 80799 Munich, Germany.}

\date{\today}

\begin{abstract}

    Optical antennas link objects to light. Here, we analyze metal
    nanorod antennas as cavities with variable reflection coefficients
    to derive the interaction of dipolar transitions with radiation
    through the antenna modes. The presented analytical model accurately
    describes the complete emission process, and is summarized in a
    phase-matching equation. We show how antenna modes evolve as they
    become increasingly more bound, i.e. plasmonic. The results
    illustrate why efficient antennas should not be too plasmonic, and
    how subradiant even modes can evolve into weakly-interacting dark
    modes. Our description is valid for the interaction of nanorods
    with light in general, and is thus widely applicable.

\end{abstract}

\maketitle

Optical antennas improve the interaction of an object with optical
radiation by means of a near-field coupling. The object absorbs and
emits light through the antenna modes
\cite{Greffet1,TaminiauNatPhot}. Metallic nano-particles are
especially suited as optical antennas because they support confined
plasmon modes that respond strongly to light \cite{Crozier1,
Muhlschlegel}. With optical antennas, the electronic transitions of
quantum emitters, such as molecules and quantum dots, can be
controlled. Excitation and emission rates are enhanced \cite{Kuhn1,
Anger1, TaminiauNanoLett2007}, the spectral dependence shaped
\cite{Ringler1}, and the angular emission directed
\cite{TaminiauNatPhot, TaminiauOptExpr2008}.

To understand optical antennas, and how they differ from
conventional antennas, the Mie solutions are available for
ellipsoids \cite{Ringler1}, and extensive numerical studies are
performed for other shapes \cite{Bryant}. More intuitively, antennas
have been described as resonators or (Fabry-P\'{e}rot) cavities
\cite{Greffet1, Ditlbacher, DellaValle08, Bozhevolnyi1, Barnard,
Douillard1, Dorfmuller, Kolesov}. If the wave vector in/along the
cavity is known, the position of the resonant modes can be
determined \cite{Chang1,Novotny}. However, the functionality of an
antenna is not given by the value of the resonance length or
wavelength alone, but by how its modes interact with a local object
and with radiation.

In this letter, we derive the interaction of dipolar transitions
with radiation through optical antenna modes by treating the antenna
as a cavity resonator. The obtained analytical model accurately
describes the emission characteristics: the radiative decay rate,
quantum efficiency and angular emission. We use the model to study
the continuous evolution of the antenna modes from
perfectly-conducting antenna theory to quasi-static plasmonics.

        \begin{figure}[t]
        \includegraphics[scale=0.7]{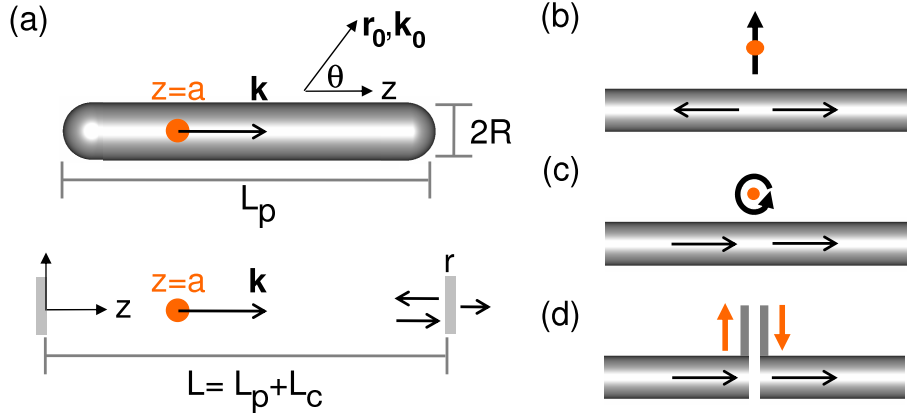}
        \caption{\label{Overview and Sources} (a) The antenna (total length $L_p$) is a rod of constant
        cross-sectional shape and size that supports a charge-density wave (wave vector $k$), and is terminated at
        both ends. It is driven by a local source at
        position $z=a$. We model the antenna as a 1D cavity with length $L$ and amplitude reflection coefficient
        $r$. In a comparison with numerical calculations, we study cylindrical antennas
        with radius $R$ and hemispherical ends. (b)-(d)
        Three local sources and the direction of the waves induced: (b) electric and (c) magnetic dipole, (d) transmission line}
        \end{figure}

We consider an elongated antenna of physical length $L_{p}$ with a
central section, of constant cross-sectional shape and size, that
supports a charge density wave with wave vector:

\begin{equation}
k = k'+ik''.
\end{equation}

The antenna is terminated at both ends, forming a resonator, which
we model as a two-mirror cavity, Fig. \ref{Overview and Sources}(a).
The model developed applies to any cross-sectional shape, assuming
$k$ is known.

The waves originate from a source at position $a$ along the antenna
axis. Figure \ref{Overview and Sources} shows three different
sources: an electric dipole, a magnetic dipole and a transmission
line. The dipoles represent electronic dipolar transitions;
transition rates are proportional to the emitted power. Electric and
magnetic dipoles differ in the direction of the induced waves on
opposite sides of the source. The transition rate for an electric
(magnetic) dipole depends on the electric (magnetic) mode density,
i.e. the magnitude of the impedance. The transmission line resembles
a magnetic dipole, but the fraction of the energy fed into the
antenna is determined by impedance matching instead \cite{Alu1,
Alu2, Burke1}. As a result, dipolar transitions dominantly excite
different modes than standard center-fed antennas.

Resonant modes are expected for physical antenna lengths that are
shifted from the multiples of $\pi/k'$ by a constant value
\cite{TaminiauNanoLett2007, Novotny, Dorfmuller}. When modeling the
antenna as a cavity, this displacement can be introduced by a
positive phase shift upon reflection \cite{DellaValle08, Barnard,
Dorfmuller} or by an extended cavity length \cite{Novotny, Alu1}.
The two corrections give the same resonant length, but are otherwise
not equivalent. We choose to set an extended length $L=L_p+L_c$ and
a real-valued reflection coefficient $r$.

To derive the resultant current distribution $I(z,a)$ we do not
distinguish between conduction and displacement currents and assume
a one-dimensional (1D) sinusoidal distribution. A superposition in
complex notation for time-harmonic waves gives, for $0\leqq z<a$,

    \begin{equation}
    I(z,a) = \frac{I_0(re^{ikz}-e^{-ikz})}{1-r^2e^{2ikL}}(e^{ika}\pm
    re^{-ika}e^{2ikL}),
    \label{Current1}
    \end{equation}
and, for $a<z\leqq L$,
    \begin{equation}
    I(z,a) = \frac{I_0(re^{ika}\pm
    e^{-ika})}{1-r^2e^{2ikL}}(e^{ikz}-re^{-ikz}e^{2ikL}).
    \label{Current2}
    \end{equation}
The initial amplitude of the induced wave $I_0$ depends on the type
of dipole, its oscillator strength, and the three-dimensional (3D)
configuration.

The $+$ signs in equations \ref{Current1} and \ref{Current2} are for
electric dipoles, the $-$ signs for magnetic dipoles; electric and
magnetic dipoles couple effectively to the antenna modes at
different positions, a result of the geometrical argument in figure
\ref{Overview and Sources}. The magnetic mode density maxima
coincide with the electric mode density minima.

The diffracted far-field observed at $r_0$ is given by:

    \begin{equation}
    E_\theta =
    i\eta_0\frac{k_0e^{ik_0r_0}}{4{\pi}r_0}\sin\theta\int_{0}^{L}I(z,a)e^{-ik_0z\cos\theta}dz,
    \label{Diffraction integral}
    \end{equation}
in which $\eta_0$ is the impedance and $k_0$ the wave vector for the
surrounding medium. The other components of the electric field are
zero. After evaluating the integral, equation \ref{Diffraction
integral} becomes:

\begin{equation}
    E_\theta =
    \frac{iI_0E_0}{1-r^{2}e^{2ikL}}
    \Big [A\big (\frac{re^{-i(k_{\|}-k)a}-r}{k_{\|}-k}-\frac{e^{-i(k_{\|}+k)a}-1}{k_{\|}+k}\big )\nonumber
    \end{equation}

    \begin{equation}
    +B\big (\frac{e^{-i(k_{\|}-k)L}-e^{-i(k_{\|}-k)a}}{k_{\|}-k}-\frac{e^{-i(k_{\|}+k)L}-e^{-i(k_{\|}+k)a}}{r^{-1}e^{-2ikL}(k_{\|}+k)}\big )\Big
    ],
    \label{Angular Emission}
    \end{equation}
in which, $k_{\|}=k_0\cos(\theta)$ is the projection of $k_0$ along
the antenna axis, and
$E_0=i\eta_0k_0e^{ik_0r_0}\sin\theta/(4{\pi}r_0)$. $A=e^{ika}\pm
re^{-ika}e^{2ikL}$ and $B=re^{ika}\pm e^{-ika}$ contain the
dependence on the dipole position.

The angular emission in equation \ref{Angular Emission} is one of
the main results of this letter. It gives a complete description of
the interaction of the antenna with a dipole and with radiation. It
describes the emission of the dipole through the antenna mode and,
by reciprocity \cite{TaminiauOptExpr2008}, its excitation by
radiation.

    \begin{figure}[t]
    \includegraphics[scale=0.40]{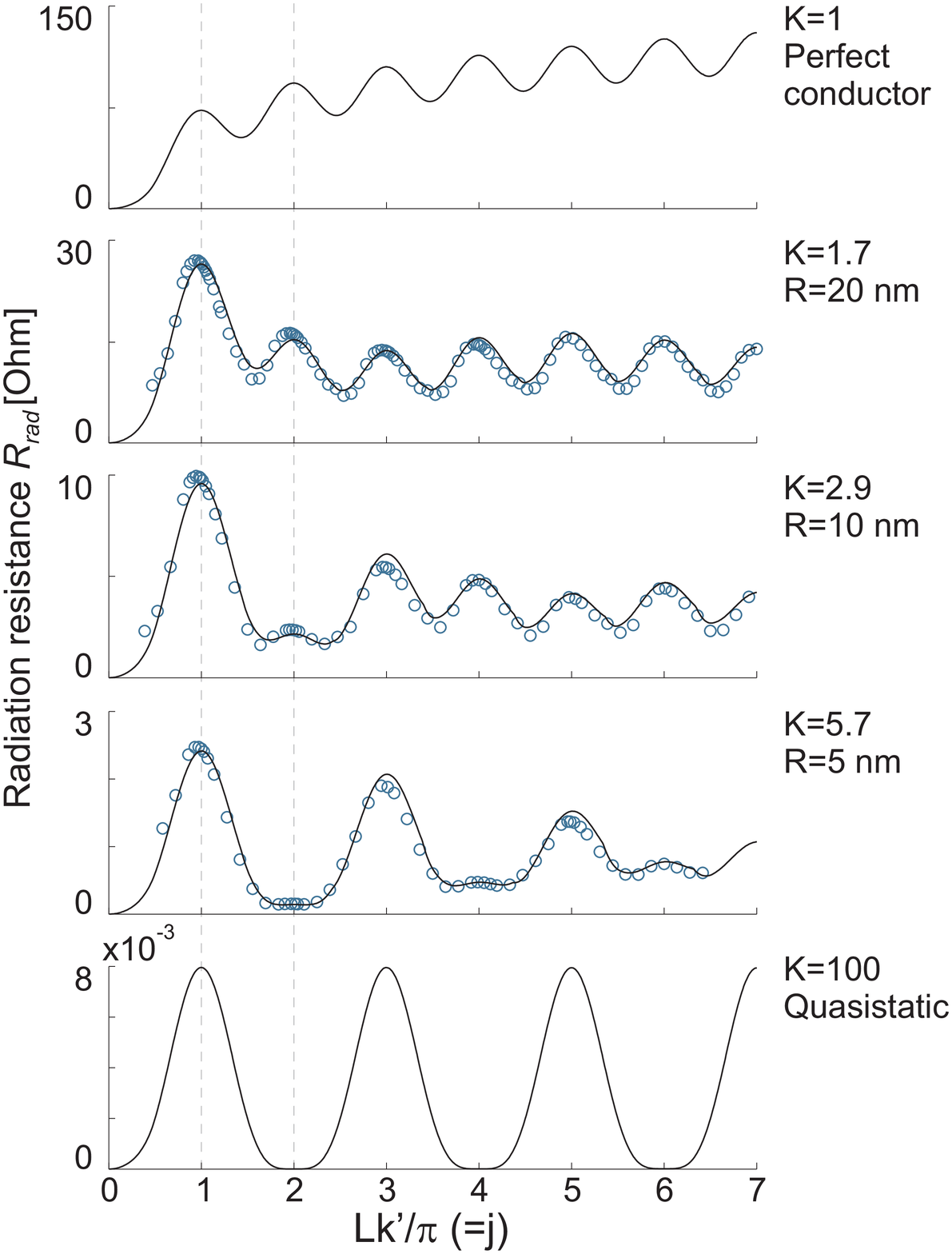}
    \caption{\label{Figure2}
    Evolution of the radiation resistance $R_{rad}(L)$ for increasingly bound antennas, i.e. increasing $K$. The optical
    antennas ($K=1.7, 2.9$ and $5.9$)
    are intermediate cases between the limits of perfect
    electrical conductors ($K=1$) and quasistatics
    ($K=100>>1$). Resonant modes occur if $Lk'/\pi=j$, with $j$ an
    integer. Lines: 1D model. Circles: 3D Numerical calculations for cylindrical gold antennas in
    vacuum (CST MicroWave Studio), Fig. \ref{Overview and
    Sources}. \textbf{Parameters} $\lambda_0=826.6$ nm. 3D Numerical: $\varepsilon_{au} =-29+2.0i$.
    Electric dipole at $5$, $2.5$ and $1.25$ nm from the antenna end,
    placed on and oriented along the antenna axis.
    1D model: Electric dipole at $a=0$ and $r=1$.
    For $K=1$ and $K=100$, $k''=0$. For $R=20, 10$ and $5$ nm: $k/k_0=1.7+0.048i, 2.9+0.11i$ and
    $5.7+0.26i$, and $L_c=54, 26$ and $12$ nm.}
    \end{figure}

Next, we use the derived model to study the main characteristics of
optical antennas in a set of concrete examples, and compare the
results to numerical simulations. We show how these characteristics
evolve as the antenna modes become increasingly more bound, i.e.
plasmonic. As a measure of how bound antenna modes are, we define an
effective index $K$:

\begin{equation}
K\equiv k'/k_0.
\end{equation}

As a concrete case we choose cylindrical gold antennas with
hemispherical ends, Fig. \ref{Overview and Sources}(a). We study
three radii, $R= 20, 10$ and $5$ nm, which give different values for
$K$ \cite{Novotny, Chang1}. We vary the antenna length $L$ for a
constant wavelength. As a source, we choose an electric dipole at
the antenna end, because it effectively excites all relevant
resonant modes: $a=0$ ($L_c$ is added right of the dipole).

To study the radiation damping, we define a radiation resistance for
$r=1$ as:

\begin{equation}
R_{rad} \equiv 2P/I_{max}^2,
\end{equation}
with $P$ the total emitted power and $I_{max}$ the maximum of
$|I(z)|$ (Eqs. \ref{Current1} and \ref{Current2}). The radiation
resistance gives the radiation damping per unit amplitude in the
resonator; it is independent of the total amplitude and is a
characteristic of the spatial distribution of the mode.

The evolution of the radiation resistance with increasingly bound
modes is illustrated in figure \ref{Figure2}, which shows $R_{rad}$
as a function of $L$ for the three optical antennas, together with
the limiting cases of $K=1$ (thin perfectly-conducting antenna), and
large $K$ (quasi-static limit). We make the following four
observations. First, the analytical results match the 3D numerical
calculations. Second, the modes excited by electric dipoles at $a=0$
differ from transmission-line center-fed antenna modes
\cite{Balanis}. Magnetic dipoles at $a=L/2$ do reproduce the results
for center-fed perfectly-conducting \cite{Balanis} ($K=1$) and
carbon nano-tube \cite{Burke1} ($K=100$) antennas. Third, unlike for
$K=1$, the radiation resistance for optical antennas does not
increase with increasing length; the waves are bound. Fourth, the
radiation resistance decreases with increasingly bound modes, i.e.
increasing $K$.

In the limit of $K>>1$, equation \ref{Angular Emission} yields
$R_{rad}\propto 1/K^2$, and since $K\propto1/R$ \cite{Chang1}, this
also implies $R_{rad}\propto R^2$. The resonances are scale
invariant only if $L_c \propto R$, or equivalently if the reflection
phase is constant, which explains previous unexpected calculation
results \cite{Kolesov}, and definitions \cite{Novotny}.

We label the resonant modes $j=1,2...$ with $L=j\pi/k'$. Even and
odd modes evolve differently; the radiation resistance of even modes
diminishes with increasing $K$. These modes have anti-symmetric
current distributions and no net dipole moment. For small antennas,
i.e. $K>>1$, these modes become subradient; the antenna modes
scatter in all directions, opposite-oriented current elements cancel
and the radiation resistance tends to zero. The nanorod antenna is
not a simple Fabry-P\'{e}rot cavity. The dependence of the radiation
resistance on the antenna length $L$ implies that the reflection
coefficient $r$ is not a constant. We relate $r$ to the radiation
resistance:

    \begin{equation}
    r(L) = \frac{Z-R_{rad}/2}{Z+R_{rad}/2},
    \label{reflection coefficient}
    \end{equation}
in which $Z$ is the real part of the antenna wave impedance, which
is taken as a free parameter here. Equation \ref{reflection
coefficient} is obtained by equating the reflection loss in the
cavity model with the antenna radiation. It assumes: that all
radiative loss is due to reflection, which is partly justified for
bound waves; that the dissipation is small so that $I_{max}$ is an
approximate measure for the current at all positions; and that
$R_{rad}$ depends weakly on $r$.

    \begin{figure}[t]
    \includegraphics[scale=0.40]{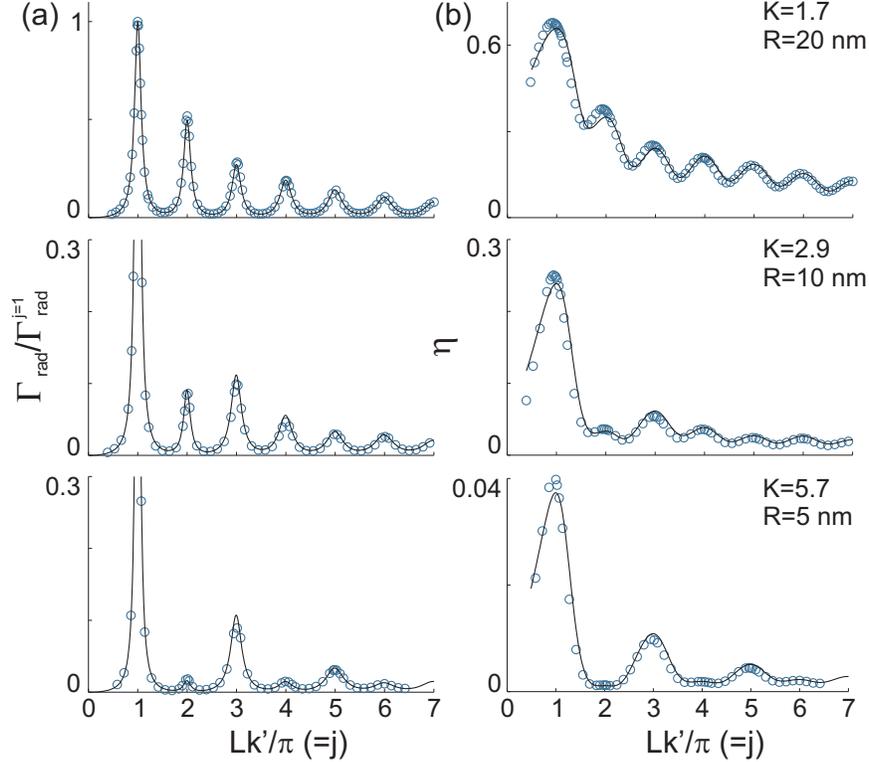}
    \caption{\label{Figure3}
    The radiative transition rate $\Gamma_{rad}$ relative to the rate for $j=1$ (a) and the quantum efficiency
    $\eta$ (b)
    for the three optical antennas. All parameters as in figure 2, but $r$ from equation \ref{reflection coefficient} with $Z=165, 250$ and
    $450 \Omega$.}
    \end{figure}

With $r$ defined, the radiative transition rate $\Gamma_{rad}$
($\propto P$) can be compared quantitatively for the different
resonant modes, Fig. \ref{Figure3}(a). The relative values for
$\Gamma_{rad}$ agree well with the numerical results. If $r$ is
taken constant instead, larger deviations are observed (e.g. for
$R=10$ nm, the error for the $j=2$ peak is $8\%$ for $r(L)$ and
$18\%$ for $r$ constant). While for $K=1$ all modes are pronounced,
even modes disappear with increasing $K$. If the loss is dominated
by dissipation, subradiant modes evolve into dark modes with small
$\Gamma_{rad}$, despite the decreased damping due to low $R_{rad}$.
By reciprocity, a small $\Gamma_{rad}$ implies low field
enhancements \cite{TaminiauOptExpr2008}; these dark modes interact
weakly with radiation. For larger $K$, higher order modes are weaker
compared to the $j=1$ mode; for low $R_{rad}$ the dissipative losses
($k''$, per length) dominate the radiative losses ($r$, per
reflection/roundtrip), and $\Gamma_{rad}$ decays quickly with
increasing $L$.

The balance between radiative and dissipative ($\Gamma_{nr}$) rates
gives rise to a quantum efficiency
$\eta=\Gamma_{rad}/(\Gamma_{rad}+\Gamma_{nr})$. The intrinsic
efficiency of the dipole emitter is taken as unity. The constant
additional dissipation due to the proximity of the dipole to the
metal is not included in the model and is subtracted from the
numerical results. Thus, $\eta$ is the antenna efficiency and sets
an upper limit to the quantum efficiency of emission through the
antenna modes.

The efficiency generally decreases with $K$, particularly for the
even modes, because the radiation resistance decreases, Fig.
\ref{Figure3}(b). Clearly, an efficient antenna should not be too
plasmonic and should in general operate away from the quasi-static
small-particle plasmon resonance. In applications where efficient
conversion into a photon is not required, large-$K$ subradiant modes
with low radiation damping, and thus narrow line-widths, can be
advantageous. Examples are sensors \cite{Hao} and spasers
\cite{Bergman}.

\begin{figure}[t]
        \includegraphics[scale=0.40]{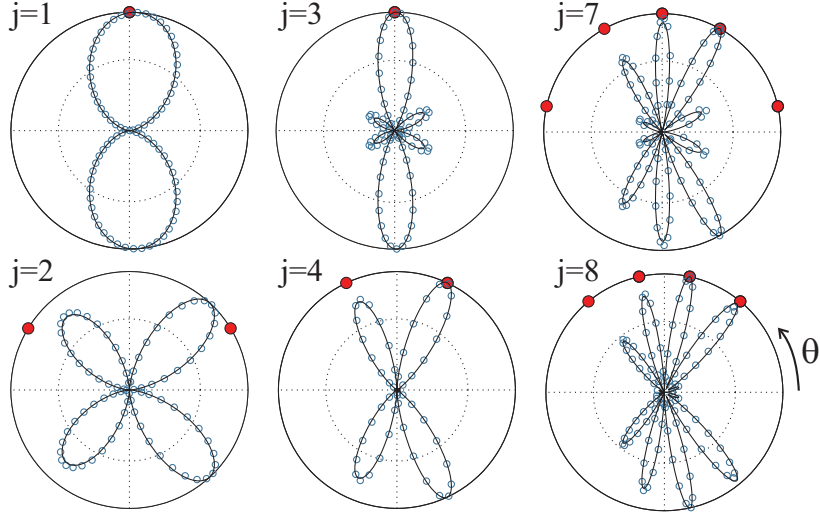}
        \caption{Angular ($\theta$) emitted power for modes $j$. $R=20$ nm, all other parameters as in figure
        \ref{Figure3}. 1D model: line. 3D numerical: circles (blue).
        Phase matching, Eq. \ref{phasematching}: dots (red).
        \label{Fig:Angular}}
        \end{figure}

The angular emission (Eq. \ref{Angular Emission}) describes under
which angles the antenna emits and can be effectively excited.
Unlike previous 0D models \cite{Encina1}, our model gives the
emission patterns of higher order modes in good agreement with
numerical calculations (Fig. \ref{Fig:Angular}). Even modes do not
interact with radiation perpendicular to the antenna axis, as
expected by symmetry arguments \cite{Schider1,Douillard1}. The lossy
nature of the modes introduces asymmetry, which reveals the position
of the dipole. Higher order modes can give multi-lobed patterns with
an odd or even amount of maxima for odd or even modes respectively.
If $K>>1$ then $k_{\|}\pm k \approx \pm k$, and the $\theta$
dependence of the denominator terms can be neglected. The emission
is then a sum of three dipole terms: $E_{0}$, $E_{0}e^{-ik_{\|}L}$
and $E_{0}e^{-ik_{\|}a}$, with the latter contribution negligible
for strong modes. The emission is given by two dipoles at the
antenna ends, making a nanorod similar to a 2-slit configuration,
and giving a basis for the intuitive picture of scattering of the
mode at the antenna ends \cite{lukin1,
Ghenuche1,Ditlbacher,Kolesov}.

We summarize the interaction of the modes with radiation in a
phase-matching equation for nano-particles:

    \begin{equation}
    k_{\|} +(2m+1)k_L=k'.
    \label{phasematching}
    \end{equation}
In which $k_L=\pi/L$, and $m=0,1,2...$ approximately give the maxima
of interaction (Fig. \ref{Fig:Angular}). Modes that do not give
solutions for equation \ref{phasematching} do not interact
effectively with radiation under any angle, and are subradiant/dark
modes. Odd modes always give at least one solution, $m=(j-1)/2$.
Even modes only yield solutions if $Lk_0>\pi$, a condition depending
only on $L$ and $k_0$ as expected for a diffraction problem.

To conclude, the derived model accurately describes the interaction
of dipolar emitters with radiation through nano-rod modes. The
antenna properties are primarily governed by a single parameter
$K=k'/k_0$ that describes how plasmonic the antenna modes are, and
are summarized in a phase-matching equation. Although here we
focused on the evolution of the emission properties for increasingly
bound waves, the model applies to all interactions with any
spatio-temporal beam and is equally valid for field enhancement and
scattering problems. The results are thus widely applicable and
might lead to further insights and design rules for optical
antennas, nano-rod spasers \cite{Bergman}, and generally for
coupling light in/from nano-rods
\cite{lukin1,Ditlbacher,Dorfmuller,Schider1, Kolesov}.

THT thanks R. Gordon for discussions.


\bibliographystyle{apsrev4-1}

%

\end{document}